\documentclass[journal]{IEEEtran}
\usepackage{cite}
\usepackage{amsmath,amssymb,amsfonts}
\usepackage{graphicx}
\usepackage{textcomp}
\usepackage{xcolor}
\usepackage{float}
\usepackage{placeins}
\usepackage{tikz}

\usepackage{graphicx}

\usepackage{algorithm}
\usepackage{algpseudocode}

\usepackage{tabularx}
\usepackage{siunitx}
\usepackage{multirow}
\usepackage{booktabs}

\usetikzlibrary{matrix,positioning,arrows.meta,shapes,fit,backgrounds,calc}

\usepackage[caption=false,font=footnotesize]{subfig}
\usepackage[utf8]{inputenc}
\algrenewcommand\algorithmiccomment[1]{\hfill// #1}

\def\BibTeX{{\rm B\kern-.05em{\sc i\kern-.025em b}\kern-.08em
    T\kern-.1667em\lower.7ex\hbox{E}\kern-.125emX}}


\usepackage{microtype}
\microtypesetup{protrusion=true, expansion=true}
\usepackage{xurl} 
\usepackage[hidelinks]{hyperref}
\begin{document}

\title{Benchmarking Dataset for Presence-Only Passive Reconnaissance in Wireless Smart-Grid Communications}

\author{Bochra~Al~Agha and Razane~Tajeddine%
\thanks{Bochra Al Agha and Razane Tajeddine are with the Department of Electrical and Computer Engineering, American University of Beirut, Beirut, Lebanon (e-mail: baa76@mail.aub.edu; razane.tajeddine@aub.edu.lb).}%
}

\maketitle

\begin{abstract}
Benchmarking presence-only passive reconnaissance in smart-grid communications is challenging because the adversary is receive-only, yet nearby observers can still alter propagation through additional shadowing and multipath that reshapes channel coherence. Public smart-grid cybersecurity datasets largely target active protocol- or measurement-layer attacks and rarely provide propagation-driven observables with tiered topology context, which limits reproducible evaluation under strictly passive threat models. This paper introduces an IEEE-inspired, literature-anchored benchmark dataset generator for passive reconnaissance over a tiered Home Area Network (HAN), Neighborhood Area Network (NAN), and Wide Area Network (WAN) communication graph with heterogeneous wireless and wireline links. Node-level time series are produced through a physically consistent channel-to-metrics mapping where channel state information (CSI) is represented via measurement-realistic amplitude and phase proxies that drive inferred signal-to-noise ratio (SNR), packet error behavior, and delay dynamics. Passive attacks are modeled only as windowed excess attenuation and coherence degradation with increased channel innovation, so reliability and latency deviations emerge through the same causal mapping without labels or feature shortcuts. The release provides split-independent realizations with burn-in removal, strictly causal temporal descriptors, adjacency-weighted neighbor aggregates and deviation features, and federated-ready per-node train, validation, and test partitions with train-only normalization metadata. Baseline federated experiments highlight technology-dependent detectability and enable standardized benchmarking of graph-temporal and federated detectors for passive reconnaissance.
\end{abstract}

\section{Introduction}

Modern smart grids rely on heterogeneous, multi-layer communication infrastructures that interconnect field devices, substations, control centers, and wide-area monitoring systems. These communication layers---typically categorized into the Home Area Network (HAN), Neighborhood Area Network (NAN), and Wide Area Network (WAN)---integrate diverse technologies (e.g., IEEE~802.15.4/ZigBee-class links, Wi-Fi, Power-Line Communication (PLC), LTE, Ethernet, and optical fiber) within interoperability frameworks and reference architectures aligned with IEEE smart-grid guidance (e.g., IEEE~2030 and IEEE~2030.5) \cite{IEEE2030-2011,IEEE2030_5,gungor2011commtech}. As the penetration of distributed energy resources (DERs), smart meters, and automated demand-response expands, the communication surface becomes increasingly wireless, distributed, and exposed to adversarial observation.

Smart-grid cybersecurity research has extensively examined active attacks, including false-data injection, replay, route manipulation, and denial-of-service \cite{yan2012survey,kim2023smart}. Passive reconnaissance threats remain comparatively under-explored. This paper focuses on presence-only passive reconnaissance: a receive-only adversary (no injection, replay, jamming, or message manipulation) that is physically proximate to a link or endpoint, so the surrounding propagation environment is measurably perturbed through additional shadowing and multipath variation \cite{ngamakeur2020survey}. Such effects are well established in device-free Radio Frequency (RF) sensing and localization, where human or object presence perturbs Received Signal Strength (RSS) and Channel State Information (CSI) statistics without protocol-layer tampering \cite{wilson2010radio,palipana2016channel}. In smart-grid settings, these perturbations manifest as low-amplitude, temporally correlated deviations in link indicators (e.g., CSI amplitude, Signal-to-Noise Ratio (SNR), packet error, and latency), motivating detection methods that go beyond static thresholding.

This work is intentionally framed as a controlled synthetic benchmark for propagation-level perturbations under a presence-only threat model, rather than a full network emulator or a field-calibrated digital twin. Public benchmarks capturing receive-only, proximity-induced channel perturbations in a multi-node HAN--NAN--WAN topology with federated-ready per-node partitions remain scarce; consequently, synthetic generation is a practical route to support leak-safe, reproducible benchmarking for topology-aware detection.

\smallskip
\noindent\textbf{Contributions:}
\begin{itemize}
    \item \textbf{Topology-aware benchmark:} a 12-node HAN--NAN--WAN communication graph with role/layer/technology assignments and tier-aware adjacency constraints (no direct HAN--WAN links).
    \item \textbf{Strictly passive perturbations:} receive-only attacks modeled as propagation changes (shadowing and coherence degradation), with link indicators recomputed coherently via measured-CSI $\rightarrow$ SNR $\rightarrow$ Packet Error Rate (PER) $\rightarrow$ latency (no injection, replay, modification, or event flags).
    \item \textbf{Leak-safe construction:} split-independent realizations with burn-in removal, strictly causal rolling features, and train-only per-node standardization applied to validation/test via stored parameters.
    \item \textbf{Temporal + neighbor context:} causal time-series descriptors augmented with adjacency-weighted neighbor aggregates and deviation features to support interpretable topology-aware learning.
    \item \textbf{Federated-ready release:} per-node train/validation/test partitions with node-specific normalization metadata, plus exported topology and node metadata for centralized, local, and federated graph-temporal pipelines.
\end{itemize}

\section{Related Work}
\label{sec:related_work}

Public smart-grid cybersecurity datasets predominantly emphasize \emph{active} attacks and higher-layer artifacts---e.g., false-data injection, denial-of-service, spoofing, and delay/drop anomalies---captured through power-system measurements, protocol traces, or flow-level features \cite{alani2023survey,tan2024high}. These releases are valuable for intrusion characterization under operationally realistic supervisory control settings, yet they rarely expose \emph{propagation-layer} observables (e.g., CSI/RSSI amplitude/phase proxies, shadowing/interference dynamics, or coherence-related statistics) that can drift under receive-only proximity.

IEC~61850-oriented benchmarks and generators further concentrate on protocol and traffic signatures (e.g., GOOSE/MMS behavior, message semantics, timing, and engineered packet/flow features), which are complementary to but distinct from passive, propagation-only effects \cite{quincozes2023ereno}. Consequently, the majority of available smart-grid benchmarks support detection of protocol-layer intrusions and active disruptions, rather than subtle channel-induced deviations that arise without packet injection or modification.

Outside the smart-grid domain, widely used intrusion datasets such as CIC-IDS2017 provide standardized, flow-based baselines but omit grid-specific semantics (HAN/NAN/WAN tiering, role-driven periodicity, and technology heterogeneity) that shape realistic smart-grid communication surfaces \cite{sharafaldin2018toward}. Similarly, wireless intrusion datasets and surveys (e.g., WSN-DS and empirical 802.11 intrusion datasets) focus on MAC/routing-layer threats and radiotap/traffic-derived indicators, without topology-aware smart-grid structure, tiered entity roles, or federated-ready per-node partitions \cite{almomani2016wsn,DBLP:journals/comsur/KoliasKSG16}. The resulting gap is a scarcity of benchmarks that jointly provide (i) tiered HAN--NAN--WAN topology context, (ii) heterogeneous technologies, and (iii) strictly passive, propagation-only perturbations suitable for leak-safe, reproducible ML evaluation.

\section{Design Goals and Threat Model}
\label{sec:design_goals_threat_model}

\subsection{Design Goals}

The generator is designed to support reproducible benchmarking of presence-only passive reconnaissance under a tiered smart-grid communication topology.

\textbf{G1) Tiered topology (IEEE-inspired):}
A HAN--NAN--WAN communication graph is instantiated with role-consistent node assignments and tier-aware connectivity constraints, enabling topology-aware modeling without direct HAN--WAN edges.

\textbf{G2) Propagation-only attacks:}
No packets are injected, replayed, modified, jammed, or dropped. Attacks are modeled strictly as propagation perturbations (shadowing and coherence degradation), and link indicators are recomputed coherently via CSI amplitude (\texttt{C}) $\rightarrow$ SNR (\texttt{SNR})  $\rightarrow$ PER \texttt{PER} $\rightarrow$ latency (\texttt{L}).

\textbf{G3) Leak-safe evaluation:}
Train/validation/test are generated as split-independent realizations with burn-in removal; engineered features are strictly causal; normalization is fit per node on train-only statistics and applied to validation/test using stored parameters.

\textbf{G4) Topology-aware learning support:}
The release includes neighbor-aware summaries derived from adjacency-weighted aggregation, enabling graph-temporal pipelines under centralized, local, or federated training.
\subsection{Threat Model}
\label{subsec:threat_model}

\textbf{Adversary class (presence-only, receive-only):}
The benchmark targets a communication-layer adversary that remains \emph{receive-only} while operating in close physical proximity to select non-wired links or endpoints. The adversary does not transmit or perform protocol-layer actions, \emph{i.e.}, no packet injection, replay, modification, jamming, routing manipulation, or intentional dropping. The only modeled mechanism is a proximity-induced perturbation of the propagation environment that is sufficiently local to alter measured link indicators on legitimate transmissions.

\textbf{Attack surface and eligible media:}
In this work, targets are restricted to non-wired media commonly used across HAN/NAN/WAN layers (e.g., IEEE~802.15.4/ZigBee and Wi-Fi in HAN; PLC and cellular backhaul in NAN/WAN), reflecting the heterogeneous nature of smart-grid communication stacks \cite{gungor2011commtech,guzelgoz2011review,IEEE802154,IEEE1901}. 
Fiber/Ethernet backbone links are designated as attack-ineligible and are kept labeled normal throughout the benchmark.

\textbf{Physical mechanism and observability:}
Receive-only proximity can still produce measurable channel perturbations because a nearby human/object and associated equipment alter absorption, diffraction, and multipath scattering in the dominant interaction region of the link. Device-free RF sensing and radio tomographic imaging demonstrate that a person carrying no device can be inferred from RSS variations on existing links \cite{wilson2010radio,wilson2012skew}, and CSI-based human presence detection shows that multipath structure changes are observable through commodity Wi-Fi physical-layer measurements \cite{palipana2016channel,zhou2015multipath}. For low-power 2.4\,GHz sensor links, human-body shadowing has been reported to introduce non-negligible additional attenuation and fading that can reduce effective SNR and increase packet loss \cite{januszkiewicz2018shadowing}. Standardized channel modeling further treats environmental changes through statistical large-scale terms (e.g., shadow fading variance and decorrelation) that can shift with geometry and blockage conditions \cite{3GPP38901}. In this benchmark, these proximity effects are expressed as (i) additional shadow-loss on active transmissions and (ii) a reduction in channel coherence (temporal correlation drop with increased innovation), producing subtle but structured deviations in channel-derived indicators.

\textbf{Capabilities and constraints:}
The modeled adversary is presence-only: it is positioned near connected groups of attack-eligible nodes during selected time windows, and the resulting proximity-driven channel variations appear as temporally correlated propagation perturbations in the measured link indicators during periods of legitimate activity. Link indicators are recomputed through a causal chain consistent with communication-system structure:
\[
\text{measured CSI/RSSI} \rightarrow \text{SNR} \rightarrow \text{PER} \rightarrow \text{latency},
\]
so that channel perturbations propagate coherently into packet reliability and delay. Detection is therefore posed on observables such as CSI amplitude/phase proxies, inferred SNR, packet error, latency (and their causal rolling descriptors), augmented by topology-based neighbor context.

\textbf{Defender objective:}
The defender’s task is a binary detection of low-amplitude, propagation-only anomalies on attack-eligible nodes using time-series features and topology-aware context derived from the HAN--NAN--WAN adjacency. The benchmark emphasizes stealth: perturbations are designed to be modest, temporally correlated, and activity-gated, avoiding trivial threshold-triggered artifacts.

\section{Topology and Feature Generation}
\label{sec:topology_features}

\subsection{Communication Topology}

The communication graph follows a tiered HAN/NAN/WAN structure aligned with IEEE smart-grid interoperability abstractions (e.g., IEEE~2030 and related profiles) \cite{basso2012ieee,IEEE2030-2011,IEEE2030_5}. Each node is assigned (i) a tier (HAN/NAN/WAN), (ii) an operational role, and (iii) a representative communication technology. The instantiated benchmark uses 12 nodes with the following layer and communication-technology assignment:

\begin{itemize}
\item HAN: SmartMeter$_0$ (ZigBee), SmartMeter$_1$ (ZigBee), SmartMeter$_2$ (Wi-Fi), Gateway$_3$ (Wi-Fi)
\item NAN: DER$_4$ (LoRa), DER$_5$ (LoRa), FeederRelay$_6$ (PLC), Controller$_7$ (LTE)
\item WAN: PMU$_8$ (Fiber), SCADA$_9$ (Fiber), AMIHeadend$_{10}$ (LTE), SubstationGW$_{11}$ (PLC)
\end{itemize}

While operational distribution systems can involve hundreds of buses and large numbers of communicating endpoints, this benchmark intentionally instantiates a compact 12-node communication graph aligned with IEEE 2030 tier abstractions (HAN/NAN/WAN). The reduced topology isolates the functional communication attack surface and enables reproducible evaluation of topology-aware and federated detectors across heterogeneous media (e.g., ZigBee, PLC, LTE) without requiring full-scale power-system emulation. The generator remains parameterized, allowing larger node inventories and adjacency structures when needed.

\begin{figure}[t]
\centering
\resizebox{\linewidth}{!}{%
\begin{tikzpicture}[
    font=\scriptsize,
    device/.style={
        draw, rounded corners, align=center, inner sep=2pt,
        minimum width=18mm, minimum height=9mm
    },
    link/.style={line width=0.5pt},
    layerbox/.style={draw, dashed, rounded corners, inner sep=8pt}
]

\node[device] (n8)  at (-3.8, 8.2) {\textbf{PMU$_8$}\\FB};
\node[device] (n9)  at (-1.3, 8.2) {\textbf{SCADA$_9$}\\FB};
\node[device] (n10) at ( 1.2, 8.2) {\textbf{AMI$_{10}$}\\LTE};
\node[device] (n11) at ( 3.7, 8.2) {\textbf{S-GW$_{11}$}\\PLC};
\node[] (nwan) at (-4.6, 9){};

\node[device] (n7) at (0.0, 5.6) {\textbf{CTRL$_7$}\\LTE};
\node[device] (n4) at (-3.0, 4.0) {\textbf{DER$_4$}\\LR};
\node[device] (n5) at ( 0.0, 4.0) {\textbf{DER$_5$}\\LR};
\node[device] (n6) at ( 3.0, 4.0) {\textbf{FR$_6$}\\PLC};

\node[device] (n3) at (0.0, 1.6) {\textbf{GW$_3$}\\WF};
\node[device] (n0) at (-3.0, 0.0) {\textbf{SM$_0$}\\ZB};
\node[device] (n1) at ( 0.0, 0.0) {\textbf{SM$_1$}\\ZB};
\node[device] (n2) at ( 3.0, 0.0) {\textbf{SM$_2$}\\WF};

\begin{pgfonlayer}{background}

\node[layerbox, fit=(nwan)(n9)(n10)(n11)] (boxWAN) {};
\node[layerbox, fit=(n4)(n5)(n6)(n7)]   (boxNAN) {};
\node[layerbox, fit=(n0)(n1)(n2)(n3)]   (boxHAN) {};

\node[anchor=north west, font=\scriptsize\bfseries]
  at ([xshift=4.5mm,yshift=-1.5mm] boxWAN.north west) {WAN};
\node[anchor=north west, font=\scriptsize\bfseries]
  at ([xshift=7mm,yshift=-3mm] boxNAN.north west) {NAN};
\node[anchor=north west, font=\scriptsize\bfseries]
  at ([xshift=7mm,yshift=-3mm] boxHAN.north west) {HAN};

\draw[link] (n8.east) -- (n9.west);
\draw[link] (n9.east) -- (n10.west);
\draw[link] (n10.east) -- (n11.west);

\draw[link] (n3.south)      -- (n1.north);
\draw[link, shorten < = -2pt] (n3.south west) -- (n0.north);
\draw[link, shorten < = -2pt] (n3.south east) -- (n2.north);

\draw[link, shorten > = -2pt] (n4.north) -- (n7.south west);
\draw[link] (n7.south) -- (n5.north);
\draw[link, shorten < = -2pt] (n7.south east) -- (n6.north);

\coordinate (ct1) at ($(n7.north west)!0.20!(n7.north east)$);
\coordinate (ct2) at ($(n7.north west)!0.40!(n7.north east)$);
\coordinate (ct3) at ($(n7.north west)!0.60!(n7.north east)$);
\coordinate (ct4) at ($(n7.north west)!0.80!(n7.north east)$);

\draw[link] (n8.south)  to[out=250,in=120,looseness=1.20] (ct1);
\draw[link] (n9.south)  to[out=250,in=110,looseness=1.20] (ct2);
\draw[link] (n10.south) to[out=260,in=70 ,looseness=1.20] (ct3);
\draw[link] (n11.south) to[out=260,in=60 ,looseness=1.20] (ct4);

\draw[link]
  (n3.north) .. controls ($(n3.north)+(-2.2,1.4)$) and ($(n7.south)+(-2.2,-1.4)$) .. (n7.south);

\end{pgfonlayer}

\node[draw, rounded corners, align=left, inner sep=4pt,
      font=\scriptsize, text width=0.92\linewidth, anchor=north] at (0, -1.25) {%
\textbf{Tech:} ZB=ZigBee,\ WF=Wi-Fi,\ LR=LoRa,\ PLC=Power Line Comm.,\ LTE=LTE,\ FB=Fiber
};

\end{tikzpicture}%
}
\caption{Layered smart-grid communication topology (HAN/NAN/WAN) with device roles and communication technologies.}
\label{fig:layered_topology}
\end{figure}
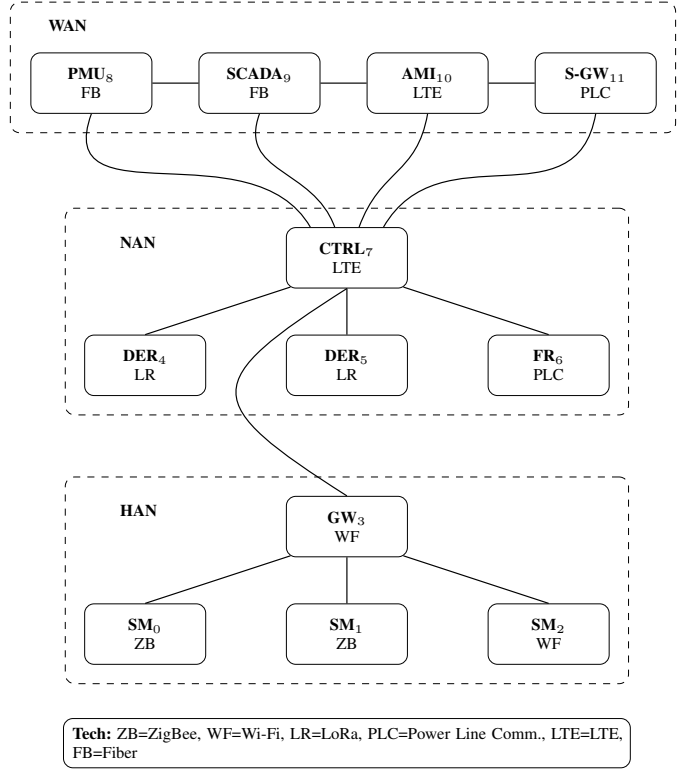

Fig.~\ref{fig:layered_topology} illustrates the HAN/NAN/WAN topology. Nodes eligible for passive-attack window placement are those using ZigBee, Wi-Fi, LTE, LoRa, or PLC; fiber nodes are excluded from attack placement in this benchmark configuration, since proximity-driven propagation perturbations primarily affect wireless/PLC links while fiber backbones are comparatively propagation-invariant.


\subsection{Topology Construction and Neighbor Modeling}

A tier-aware adjacency matrix is constructed using role-consistent connectivity constraints: (i) HAN smart meters connect only to the HAN gateway, (ii) NAN DER and feeder relay nodes connect to the NAN controller, (iii) the HAN gateway connects to the NAN controller to represent tier aggregation, and (iv) WAN backbone connectivity links the controller and substation gateway to SCADA, the AMI headend, and PMU nodes, with additional SCADA links to the AMI and PMU. This design excludes direct HAN--WAN edges while preserving aggregation and backbone relations consistent with smart-grid communication reference architectures \cite{IEEE2030-2011,gungor2011commtech}.

Neighbor-aware features are computed using a row-stochastic mixing matrix derived from the undirected binary adjacency. Let $N$ denote the number of nodes, and let $A \in \{0,1\}^{N\times N}$ denote the adjacency. Let $D=\mathrm{diag}(A\mathbf{1})$ be the degree matrix (nonzero by construction in the instantiated topology) and define the row-stochastic neighbor-averaging operator $P=D^{-1}A$. A self--neighbor convex combination is then defined as
\begin{equation}
W=\alpha I+(1-\alpha)P,\quad \alpha=0.30,
\end{equation}
which is row-stochastic. For a node-level signal $x(t)\in\mathbb{R}^N$, the neighbor aggregate is $\bar{x}(t)=Wx(t)$ and the corresponding deviation feature is $|x_i(t)-\bar{x}_i(t)|$.

\subsection{Feature Generation Pipeline}
\label{subsec:feature_pipeline}

The benchmark exports communication-layer time series generated by a causal traffic--channel--link-indicator pipeline. Node activity is represented by an integer transmission count $\mathrm{tx\_count}_i(t)$ with role-consistent dynamics (e.g., periodic metering, polling-driven supervisory traffic, near-continuous PMU telemetry, and intermittent DER/relay activity) \cite{gungor2011commtech,gomez2013smart}. 

A latent complex fading process $h_i(t)$ is generated using a technology-conditioned Gauss--Markov model (i.e., a first-order autoregressive complex Gaussian process with correlation $\rho_i$; see Sec.~\ref{Gauss Markov}), and channel strength is represented by the envelope $|h_i(t)|$ \cite{goldsmith2005wireless}. Large-scale shadowing follows standardized log-normal statistics and decorrelation behavior anchored to 3GPP TR~38.901 (Third Generation Partnership Project, Technical Report 38.901), with shared environment correlation introduced through global and layer-level components \cite{3GPP38901}. Technology-dependent interference includes correlated background terms and impulsive components (notably for PLC), consistent with canonical PLC impairment modeling \cite{katayama2006noise,zimmermann2002multipath}.

Measured amplitude $\mathrm{C}_i(t)$ is obtained from $|h_i(t)|$ with bounded observation noise and device-like quantization (IEEE~802.15.4-class links use 1\,dB quantization and bounded error) \cite{IEEE802154}. Link indicators are then recomputed coherently through the chain $\mathrm{C}\rightarrow$ SNR $\rightarrow$ PER $\rightarrow$ latency, where latency combines a technology baseline, an ARQ-inspired retransmission expectation derived from PER, jitter, and a burst component whose occurrence increases with PER; an EWMA-smoothed latency is exported to capture short-term persistence \cite{nguyen2011modeling}.

Within each split, strictly causal engineered features are computed from observables (e.g., rolling entropy/drift and moment statistics, change and activity descriptors), augmented with adjacency-weighted neighbor aggregates and deviation features. Train/validation/test are generated as independent realizations with burn-in removal, and per-node standardization is fit on train-only statistics and applied to validation/test via stored parameters.

\section{Channel Modeling and Traffic Semantics}
\label{sec:channel_model}

A technology-conditioned, temporally correlated surrogate channel is used to generate link-layer observables through a coherent chain:
\texttt{C} $\rightarrow$ \texttt{SNR} $\rightarrow$ \texttt{PER} $\rightarrow$ \texttt{L} $\rightarrow$ \texttt{$\tilde{L}$}.
The generator is designed for benchmarking communication-layer effects under heterogeneous HAN/NAN/WAN technologies rather than emulating a specific deployment.

\paragraph{Sampling semantics.}
Time is indexed by discrete epochs $t\in\{0,\ldots,T-1\}$ with duration $\Delta t$ (configured as \texttt{DT\_SECONDS} in \texttt{config.json}, and set to $\Delta t=1$\,s in the released configuration).
For node $i$, \texttt{tx\_count}$_i(t)$ represents the number of transmission attempts during epoch $t$ (possibly zero).
The exported link indicators (\texttt{C}, \texttt{SNR}, \texttt{PER}, \texttt{L}) are interpreted as an epoch-level link-quality snapshot that would govern packets attempted within that epoch.
Attack labels are activity-gated: \texttt{attack\_label}$=1$ is assigned only on epochs with \texttt{tx\_count}$>0$ for any attack-eligible (non-fiber) technology; fiber links are always labeled normal.

\subsection{Technology-Conditioned Surrogate Channel}
\label{subsec:channel_modeling}

\paragraph{Latent fading and phase.}
Each node is assigned a complex Gauss--Markov (AR(1)) fading process $h_i(t)$ with technology-dependent temporal correlation.
The complex sequence is \emph{not} exported; instead, the dataset exports (i) amplitude-derived observables via \texttt{C}, and (ii) \emph{label-safe phase descriptors} derived from $h_i(t)$: \texttt{phase\_sin}, \texttt{phase\_cos}, and \texttt{dphase}. These capture short-term coherence changes commonly reflected in CSI measurements \cite{halperin2011tool}.

\paragraph{Shadowing and interference.}
Large-scale fading is modeled in dB as a temporally correlated shadowing process with scenario-dependent standard deviations and decorrelation distances anchored to 3GPP TR~38.901 (e.g., InH/UMi/UMa categories) \cite{3GPP38901}.
To reflect shared environmental conditions, shadowing and interference are composed from global, layer-level, and node-local components, and an optional correlation template is applied following the classical exponentially decaying shadowing-correlation model \cite{gudmundson1991correlation}.
In addition, \texttt{interf\_db} includes technology-specific burstiness; for PLC, impulsive components are modeled using standard impulsive-noise abstractions (mixture/Markov-like and heavy-tailed impulses) consistent with PLC noise literature \cite{middleton1977statistical,katayama2006noise,zimmermann2002multipath}.

\paragraph{Measurement model.}
A measured amplitude proxy \texttt{C} is produced by mapping the latent amplitude to dB, adding bounded observation noise, and quantizing to device-like resolution.
For ZigBee nodes, the RSSI-like measurement follows 1\,dB quantization and bounded multi-dB accuracy consistent with CC2420-class radios \cite{cc2420datasheet}.
Other technologies use a lower-noise bounded measurement model.

\paragraph{Channel-to-metrics mapping.}
SNR (in dB) is computed from \texttt{C} using a log-amplitude mapping with (i) a technology-dependent base SNR, (ii) a per-node static offset capturing calibration variability, (iii) a technology-specific link margin, and (iv) additive \texttt{shadow\_db} and subtractive \texttt{interf\_db}.
Packet error rate is then derived via a technology-dependent logistic mapping.
For LoRa nodes, sensitivity behavior is anchored through spreading-factor-dependent operating points (consistent with LoRa receiver specifications) \cite{semtechsx1276datasheet}.
Latency is computed as the sum of (i) a technology-specific baseline, (ii) an ARQ-inspired retransmission expectation derived from packet error (capped for stability), (iii) Gaussian jitter, and (iv) a burst component whose occurrence rate increases with packet error; \texttt{$\tilde{L}$} is exported as an EWMA-filtered latency trace.

\paragraph{Diagnostics and feature hygiene.}
The generator exports \texttt{shadow\_db} and \texttt{interf\_db} as \emph{auxiliary diagnostics} to validate the channel-to-metrics chain and support controlled ablations.
These streams are not directly observable at the receiver and are therefore excluded from the recommended ML feature set (and blacklisted in the accompanying baselines), while the primary learning features remain derived from measurable link indicators (\texttt{C}, \texttt{SNR}, \texttt{PER}, \texttt{$\tilde{L}$}) and their causal temporal/graph aggregates.

\subsection{Physical Consistency and Statistical Alignment}
\label{subsec:physical_consistency}

Because the benchmark is synthetic, credibility relies on two properties: (i) exported variables must follow a coherent cause--effect chain, and (ii) the non-attack regime should remain consistent with standard abstractions used in wireless/PLC propagation and smart-grid communication modeling \cite{gungor2011commtech,guzelgoz2011review,zimmermann2002multipath,katayama2006noise,3GPP38901}.

\subsubsection{End-to-end coherence of exported observables}
\label{subsec:causal_chain}

The generator enforces a deterministic channel-to-metrics mapping, so downstream variables change only through their physical dependencies (no injected flags).
Presence-only perturbations modify propagation terms (e.g., shadowing and local coherence/scattering), which is consistent with device-free RF sensing evidence that nearby bodies/objects can induce measurable RSS/CSI changes without protocol manipulation \cite{januszkiewicz2018shadowing,wilson2010radio,zhou2015multipath,palipana2016channel}.
These changes propagate through the exported chain:
\texttt{C} $\rightarrow$ \texttt{SNR} $\rightarrow$ \texttt{PER} $\rightarrow$ \texttt{L} $\rightarrow$ \texttt{$\tilde{L}$},
with \texttt{C} treated as a noisy amplitude proxy and with SNR computed in the dB domain (link-budget form) \cite{halperin2011tool,rappaport2010wireless,goldsmith2005wireless}.
PER uses a technology-conditioned monotone ``waterfall'' mapping from SNR, preserving physical dependence without fixing a specific PHY/MCS table \cite{john2008digital}.
Latency is derived from PER via an ARQ-inspired retransmission proxy plus jitter and burst components, consistent with smart-grid latency abstractions \cite{nguyen2011modeling,gomez2013smart}.

\subsubsection{Normal-regime alignment}
\label{subsec:statistical_alignment}

\textbf{Small-scale variability.}
Normal fluctuations arise from a temporally correlated complex Gauss--Markov fading process, with exported amplitude and phase-derived descriptors.
This yields controllable coherence and scattering variability while remaining consistent with standard fading families \cite{rappaport2010wireless,goldsmith2005wireless}.

\textbf{Shadowing and correlation.}
Shadowing is generated as a Gaussian process in dB (lognormal in linear scale) with scenario-dependent variability and decorrelation anchored to 3GPP TR~38.901, and includes global/layer/local components to induce realistic co-movement across nodes \cite{3GPP38901,gudmundson1991correlation}.
Key numerical anchors used by the generator are summarized in Table~\ref{tab:param_anchoring}.

\textbf{PLC/noise-driven burstiness.}
Interference includes technology-dependent colored components and burst terms; for PLC, impulsive-noise abstractions are used to reproduce heavy-tailed impairment behavior commonly reported in PLC modeling \cite{middleton1977statistical,katayama2006noise,zimmermann2002multipath}.

\textbf{Traffic and delay semantics.}
Activity (\texttt{tx\_count}) follows role-conditioned background behavior with periodic components and rare bursts, matching the qualitative structure of HAN/NAN/WAN communication roles described in smart-grid communication surveys and guidelines \cite{IEEE2030-2011,NISTIR7628,gungor2011commtech}.
Delays combine baseline technology latency with PER-driven retransmission effects and bursty components, producing realistic non-Gaussian tails \cite{nguyen2011modeling,gomez2013smart}.
\subsubsection{Parameter grounding in standards, datasheets, and measurement literature}
To make the synthetic benchmark quantitatively defensible, the generator fixes a set of \emph{load-bearing} parameters directly from standards, datasheets, and measurement-driven modeling literature, and treats the remaining knobs as tunable proxies whose ranges are kept conservative. Table~\ref{tab:param_anchoring} summarizes the main numerical anchors used in the released configuration: large-scale shadowing variance and decorrelation distance from 3GPP TR~38.901; CC2420-class IEEE~802.15.4 RSSI quantization and bounded accuracy for ZigBee-like measurements; LoRa demodulation thresholds versus spreading factor from the SX1276 family datasheet; and PLC impulsive-noise model families widely used to reproduce heavy-tailed interference. For presence-only attacks, the shadow-loss mixture is centered on empirically observed excess attenuation magnitudes and includes a rare high-loss tail intended to cover compounded obstruction conditions (e.g., close-in human blockage plus dense/metal structures). These anchors ensure that both the normal regime and attack perturbations remain compatible with reported operating regimes rather than being arbitrarily chosen.
\begin{table*}[t]
\centering
\caption{Representative parameter anchoring: generator settings versus reported standards/datasheets/measurement literature.}
\label{tab:param_anchoring}
\scriptsize
\setlength{\tabcolsep}{4pt}
\begin{tabularx}{\textwidth}{@{}p{0.23\textwidth}p{0.25\textwidth}p{0.34\textwidth}p{0.14\textwidth}@{}}
\toprule
\textbf{Component / parameter} & \textbf{Generator setting (code)} & \textbf{Reported values / bounds in external sources} & \textbf{Source} \\
\midrule

TR~38.901 shadow fading $\sigma_{\mathrm{SF}}$ (dB) &
InH: 3.0/8.03 (LOS/NLOS); UMi: 4.0/7.82; UMa: 4.0/6.0; RMa: 4.0/8.0 &
Scenario-dependent $\sigma_{\mathrm{SF}}$ values specified by TR~38.901 for the corresponding environments (LOS/NLOS). &
\cite{3GPP38901} \\

TR~38.901 decorrelation distance $d_{\mathrm{cor}}$ (m) &
InH: 10/6 (LOS/NLOS); UMi: 10/13; UMa: 37/50; RMa: 37/120 &
Scenario-dependent $d_{\mathrm{cor}}$ values (meters) specified by TR~38.901 (LOS/NLOS). &
\cite{3GPP38901} \\

ZigBee RSSI quantization + bounded accuracy (CC2420-class) &
$q=1$\,dB; measurement noise modeled with $\sigma=3$\,dB and clipping to $\pm 6$\,dB (ZigBee case) &
CC2420 RSSI register uses 1\,dB LSB; datasheet reports typical RSSI accuracy $\pm 6$\,dB and linearity $\pm 3$\,dB (over the stated range). &
\cite{cc2420datasheet} \\

LoRa demodulation SNR vs spreading factor &
$\mathrm{SNR}_{50}$ (dB): SF7..SF12 = $\{-7.5,-10,-12.5,-15,-17.5,-20\}$ &
SX1276-family LoRa modem sensitivity guidance includes required SNR values per SF (including $\approx-20$\,dB for SF12). &
\cite{semtechsx1276datasheet} \\

Presence-only excess attenuation (attack shadow-loss) &
Mixture modes at 10, 15, 35, 55\,dB (clipped to [0.5, 67]\,dB), applied only on active epochs &
Body/presence-induced excess loss on 2.4\,GHz sensor links is reported in the multi-dB to tens-of-dB range; higher losses can occur under compounded obstruction/penetration conditions. &
\cite{januszkiewicz2018shadowing} \\

PLC impulsive-noise family (interference burstiness) &
Bernoulli--Gaussian Markov switching + $\alpha$-stable impulses; impulses clipped to 0--30\,dB &
Indoor PLC noise is widely modeled as impulsive/heavy-tailed; BG/Markov and $\alpha$-stable families are standard choices in the PLC literature. &
\cite{andreadou2010plcnoise,karakus2020plcnoise,han2019plcmerge} \\

\bottomrule
\end{tabularx}
\end{table*}

\begin{table}[t]
\caption{Proxy validation of large-scale fading statistics on the released dataset}
\label{tab:proxy_shadow_validation}
\centering
\begin{tabular}{lcccc}
\toprule
Tier (scenario proxy) & Node (tech) & $\hat{\sigma}_{\mathrm{SF}}$ [dB] & $\hat{\rho}_1$ & $\hat{d}_{\mathrm{cor}}$ [m]\\
\midrule
HAN (InH) & 0 (ZigBee) & 8.02 & 0.961 & 7.58 \\
NAN (UMi) & 4 (LoRa)   & 4.09 & 0.951 & 23.66 \\
WAN (UMa) & 8 (Fiber)  & 4.00 & 0.992 & 52.28 \\
\midrule
\multicolumn{5}{p{0.97\linewidth}}{\footnotesize
Train split ($\texttt{attack\_label}=0$, $\texttt{tx\_count}>0$); $\hat{d}_{\mathrm{cor}}$ from lag-1 correlation of $\texttt{shadow\_db}$.}
\\
\bottomrule
\end{tabular}
\end{table}

\textbf{Proxy validation on realized splits:}
Table~\ref{tab:proxy_shadow_validation} reports output-level statistics of $\texttt{shadow\_db}$ under non-attack active epochs.
The observed $\hat{\sigma}_{\mathrm{SF}}$ values align with the TR~38.901 variability ranges listed in Table~\ref{tab:param_anchoring} (InH/UMi/UMa), providing validation-by-proxy at the realization level (not only at the parameter level).

\subsection{Discrete-Time Channel Chain}
\label{subsec:channel_chain}

A discrete-time baseband abstraction is assumed,
\begin{equation}
y(t)=H(t)\,x(t)+n(t),
\end{equation}
where $H(t)\in\mathbb{C}$ is a latent complex channel response, $x(t)=1$ is a pilot symbol, and $n(t)$ models additive measurement noise \cite{goldsmith2005wireless,rappaport2010wireless}. 
Time is indexed by discrete epochs $t\in\{0,\ldots,T-1\}$ with duration $\Delta t$ seconds (configured as \texttt{DT\_SECONDS}), where $T$ denotes the number of epochs.

The dataset does not export $H(t)$; instead, it exports the deterministic chain
\begin{equation}
\begin{aligned}
\texttt{C} &\rightarrow \texttt{SNR} \rightarrow \texttt{PER} \\
&\rightarrow \texttt{L} \rightarrow \texttt{$\tilde{L}$}
\end{aligned}
\end{equation}
computed from split-local latent processes. Attack effects act only on latent propagation terms; link indicators are recomputed through the same mapping (no injected flags). Labels and perturbations are activity-gated: $\texttt{attack\_label}_i(t)=1$ is assigned only when $\texttt{tx\_count}_i(t)>0$ for attack-eligible technologies. Diagnostic impairment streams (\texttt{shadow\_db}, \texttt{interf\_db}) are released for reproducibility and should be excluded from learning.

\subsubsection{Latent complex fading and exported phase descriptors}
\label{Gauss Markov}
Let $N$ denote the number of nodes. For each node $i\in\{0,\ldots,N-1\}$, a technology-conditioned complex Gauss--Markov fading process is generated,
\begin{equation}
\begin{aligned}
h_i(t) &= \rho_i\,h_i(t-1)+\sqrt{1-\rho_i^2}\,w_i(t),\\
w_i(t) &\sim \mathcal{CN}(0,1).
\end{aligned}
\end{equation}
where $\rho_i\in[0,1)$ controls temporal correlation and is selected by the assigned communication technology \cite{goldsmith2005wireless}. Here $\mathcal{CN}(0,1)$ denotes a circularly symmetric complex Gaussian distribution. Phase descriptors are exported as 
\begin{align}
\texttt{phase\_sin}_i(t) &= \sin(\phi_i(t)), \\
\texttt{phase\_cos}_i(t) &= \cos(\phi_i(t)), \\
\phi_i(t) &= \angle h_i(t), \\
\texttt{dphase}_i(t) &= \phi_i^{\mathrm{unw}}(t)-\phi_i^{\mathrm{unw}}(t-1),
\end{align}
where $\phi^{\mathrm{unw}}_i(t)$ denotes the time-unwrapped phase of $\phi_i(t)$, with $\texttt{dphase}_i(0)=0$ by convention. These descriptors reflect short-term coherence changes observable in CSI measurements \cite{halperin2011tool}.

\subsubsection{CSI/RSSI-like amplitude observation (\texttt{C})}
Let $a_i(t)=|h_i(t)|$ denote the latent envelope. A minimum envelope floor $\epsilon_{\min}>0$ prevents $\log(0)$, and an output floor $\epsilon_0>0$ enforces positivity (both are fixed constants in \texttt{config.json}).
Define a latent dB-strength
\begin{equation}
h^{(\mathrm{dB})}_i(t)=20\log_{10}\big(\max(a_i(t),\epsilon_{\min})\big).
\end{equation}
Measurement noise is added in dB, clipped to a bounded range, quantized to resolution $q_i$ (in dB), then mapped back to amplitude:
\begin{gather}
e_i(t) \sim \mathcal{N}(0,\sigma_i^2),\\
\tilde{h}^{(\mathrm{dB})}_i(t)
= Q_{q_i}\!\Big(h^{(\mathrm{dB})}_i(t)
+\mathrm{clip}\big(e_i(t),[-c^{\mathrm{clip}}_i,c^{\mathrm{clip}}_i]\big)\Big),
\label{eq:csi_db_meas}\\
c_i(t)\triangleq \texttt{C}_i(t)
=\max\!\Big(10^{\tilde{h}^{(\mathrm{dB})}_i(t)/20},\,\epsilon_0\Big).
\label{eq:csi_amp}
\end{gather}
where $Q_{q_i}(u)=q_i\cdot\mathrm{round}(u/q_i)$ is a uniform dB quantizer, and $(\sigma_i,c^{\mathrm{clip}}_i,q_i)$ are technology-dependent parameters. For ZigBee nodes, $q_i=1$\,dB with bounded multi-dB error consistent with IEEE~802.15.4 / CC2420-class RSSI behavior \cite{IEEE802154,cc2420datasheet}.

\subsubsection{SNR derivation in dB (\texttt{SNR})}
Signal-to-noise ratio (SNR) is computed in the dB domain:
\begin{align}
\gamma^{(\mathrm{dB})}_i(t)&\triangleq \texttt{SNR}_i(t)\nonumber\\
&=\gamma^{(\mathrm{tech})}_{0,i}+\delta_i
+20\log_{10}\!\big(c_i(t)\big) \nonumber\\ &\quad
+m^{(\mathrm{tech})}_i 
 + s_{i,\mathrm{dB}}(t)- i_{i,\mathrm{dB}}(t),
\label{eq:snr_db}
\end{align}
where $\gamma^{(\mathrm{tech})}_{0,i}$ is a technology-conditioned base SNR (dB), $\delta_i$ is a per-node static offset (dB), and $m^{(\mathrm{tech})}_i$ is a technology-conditioned link margin (dB).
The terms $s_{i,\mathrm{dB}}(t)\equiv \texttt{shadow\_db}_i(t)$ and $i_{i,\mathrm{dB}}(t)\equiv \texttt{interf\_db}_i(t)$ are split-local shadowing and interference processes (dB), respectively. Shadowing uses scenario-dependent variance and decorrelation anchored to 3GPP TR~38.901 categories and may include exponential correlation following classical shadow-fading models \cite{3GPP38901,gudmundson1991correlation}. Interference includes correlated background and technology-specific burstiness; for PLC, impulsive-noise abstractions are used consistent with PLC impairment literature \cite{IEEE1901,middleton1977statistical,katayama2006noise,zimmermann2002multipath}.

\subsubsection{Packet error rate (\texttt{PER})}
Packet error rate (PER) is derived from SNR using a technology-conditioned logistic mapping:
\begin{equation}
\texttt{PER}_i(t)
=\left(1+\exp\!\left(k_i\big(\gamma^{(\mathrm{dB})}_i(t)-\gamma_{50,i}\big)\right)\right)^{-1},
\label{eq:per_sigmoid}
\end{equation}
where $\gamma_{50,i}$ (dB) is the technology-dependent SNR at which PER equals $0.5$, and $k_i>0$ controls transition steepness \cite{john2008digital}. For LoRa nodes, $\gamma_{50,i}$ is configured using spreading-factor-dependent operating points consistent with receiver specifications \cite{semtechsx1276datasheet}. The output is numerically clipped to $[\epsilon,1-\epsilon]$ for a fixed $\epsilon\in(0,1)$.

\subsubsection{Latency (\texttt{L}) and exponentially weighted moving average (EWMA) smoothing (\texttt{$\tilde{L}$})}

Let $p_i(t)=\min(\texttt{PER}_i(t),1-\epsilon)$ and define a bounded automatic repeat request (ARQ)-inspired retransmission proxy
\begin{equation}
\mathrm{reTX}_i(t)=\min\!\left(R_{\max},\frac{p_i(t)}{1-p_i(t)}\right),
\label{eq:retx_proxy}
\end{equation}
where $R_{\max}>0$ caps the expected retransmission contribution (a fixed constant).
The exported latency is then
\begin{align}
\texttt{L}_i(t)
&=L^{(\mathrm{tech})}_{0,i}
+\Delta_{\mathrm{rtx}}\;\mathrm{reTX}_i(t)
+\eta_i(t)+u_i(t),
\label{eq:latency_model}
\end{align}
where $L^{(\mathrm{tech})}_{0,i}$ is a technology-specific baseline latency (ms), $\Delta_{\mathrm{rtx}}>0$ scales retransmission delay (ms per unit reTX), $\eta_i(t)\sim\mathcal{N}(0,\sigma_{L,i}^2)$ is Gaussian jitter, and $u_i(t)\ge 0$ is a heavy-tailed burst term implemented via filtered shocks whose occurrence probability increases with $p_i(t)$ \cite{nguyen2011modeling,gomez2013smart}. The smoothed latency (\texttt{$\tilde{L}$}) is an EWMA:
\begin{equation}
\begin{aligned}
\texttt{$\tilde{L}$}_i(t)
&=\beta\,\texttt{$\tilde{L}$}_i(t-1) \\
&\quad +(1-\beta)\,\texttt{L}_i(t).
\end{aligned}
\label{eq:ewma_latency}
\end{equation}
with fixed $\beta\in(0,1)$ and split-local initialization (e.g., $\texttt{$\tilde{L}$}_i(0)=\texttt{L}_i(0)$).

\section{Passive Attack Generation}
\label{sec:attack_generation}

A strictly passive, presence-only reconnaissance threat is modeled at the communication layer. The adversary is receive-only (non-transmitting) and performs no jamming, packet injection, replay, modification, routing manipulation, or intentional dropping. Attacks are implemented solely as low-amplitude, temporally correlated perturbations of split-local propagation latents that (i) introduce additional shadow-loss and (ii) reduce temporal coherence while increasing scattering/innovation in the complex fading process. This abstraction follows the device-free RF sensing and CSI-based presence literature showing that nearby bodies/objects can measurably perturb RSS/CSI statistics without protocol-layer tampering \cite{wilson2010radio,wilson2012skew,palipana2016channel,zhou2015multipath,januszkiewicz2018shadowing}. The interference latent is not directly perturbed. Perturbations are ramped within each window and applied only on active epochs using the activity gate $\mathbb{I}\{\texttt{tx\_count}(t)>0\}$.

\subsection{Split-Local Window Sampling and Leak-Safety}
\label{subsec:attack_sampling}

Train/validation/test are generated as independent realizations using split-specific pseudo-random seeds and a burn-in of $B$ epochs (\texttt{BURN\_IN}) to stabilize correlated latents. For a split of length $T_{\text{split}}$ (after burn-in removal), each window is defined by a labeled interval $[s_0,s_1)$ and an embedded core interval $[t_0,t_1)\subset[s_0,s_1)$:
\begin{align}
0 \le s_0 < s_1 \le T_{\text{split}}, \qquad
t_0 &= s_0 + \ell_{\text{lead}}, \qquad
t_1 = t_0 + L ,
\end{align}
where $L$ is the core length (epochs) and $(\ell_{\text{lead}},\ell_{\text{tail}},\ell_{\text{hyst}})$ are lead/tail/hysteresis margins (epochs), corresponding to \texttt{LEAD}, \texttt{TAIL}, and \texttt{HYST}. The labeled length is
\begin{equation}
L_{\text{lab}} \triangleq s_1-s_0
= \ell_{\text{lead}} + L + \ell_{\text{tail}} + \ell_{\text{hyst}} .
\end{equation}
No latent state is shared across splits because channel, environment, and traffic processes are instantiated independently per split.

\subsection{Eligibility, Activity-Gated Labeling, and Coverage Target}
\label{subsec:eligibility_activity}

Only nodes whose communication technology lies in \texttt{ATTACK\_ELIGIBLE\_TECH} are eligible for window placement and labeling; attack-ineligible media (e.g., fiber/ethernet backbones) are excluded from placement and satisfy $\texttt{attack\_label}=0$ for all $t$, consistent with the heterogeneous smart-grid communication stack view \cite{gungor2011commtech,IEEE8023}. Labeling is activity-gated to avoid silent positives: link-quality indicators are meaningful only when transmissions occur, so perturbations and labels are applied only on epochs with $\texttt{tx\_count}>0$. For node $i$,
\begin{equation}
\begin{aligned}
a_i(t) 
&\triangleq \mathbb{I}\!\left[\texttt{tx\_count}_i(t)>0\right], \\
\texttt{attack\_label}_i(t)
&\leftarrow \texttt{attack\_label}_i(t)\,a_i(t).
\end{aligned}
\end{equation}

Coverage control is defined over active eligible rows. For an eligible node $i$ in a split,
\begin{equation}
\begin{aligned}
A_i &\triangleq \sum_{t=0}^{T_{\text{split}}-1} a_i(t),\\
Y_i &\triangleq \sum_{t=0}^{T_{\text{split}}-1}
\mathbb{I}\!\left[\texttt{attack\_label}_i(t)=1\right].
\end{aligned}
\end{equation}
A configurable target fraction $r$ (\texttt{TARGET\_ATTACK\_FRAC}) sets a per-node quota $Y_i \approx \lceil rA_i\rceil$, with minor deviations possible due to connected-group placement and overlap policy.

\subsection{Window Lengths, Overlaps, and Uniqueness}
\label{subsec:window_lengths_overlap}

Core durations are sampled as
\begin{equation}
L \sim \mathcal{U}_{\mathbb{Z}}\!\big[\texttt{WIN\_CORE\_MIN},\,\texttt{WIN\_CORE\_MAX}\big],
\end{equation}
where $\mathcal{U}_{\mathbb{Z}}[a,b]$ denotes the discrete uniform distribution over integers $k\in\mathbb{Z}$ such that $a\le k\le b$.
The labeled window length is $L_{\text{lab}}=\ell_{\text{lead}}+L+\ell_{\text{tail}}+\ell_{\text{hyst}}$.
Overlaps are permitted when \texttt{ALLOW\_OVERLAP=True}. Duplicate placements are prevented by enforcing a split-local uniqueness key
\begin{equation}
\big(\texttt{split},\,t_0,\,\texttt{sorted(nodes)}\big).
\end{equation}
If $\Delta t$ denotes the epoch duration (\texttt{DT\_SECONDS}), the core duration equals $L\,\Delta t$.

\subsection{Node Grouping and Locality}
\label{subsec:attack_grouping}

Each window targets a topology-local connected set of eligible nodes on the tiered HAN/NAN/WAN adjacency. An anchor node is sampled with probability proportional to its remaining coverage deficit, then the attacked set is expanded by traversing eligible 1-hop neighbors until reaching a group size
$k\in\{\texttt{GROUP\_MIN},\ldots,\texttt{GROUP\_MAX}\}$.
This locality constraint is consistent with tiered smart-grid communication organization and supports neighbor-context features defined on the same adjacency \cite{IEEE2030-2011,IEEE2030_5,gungor2011commtech}.

\subsection{Propagation-Only Perturbations with Ramp-Up}
\label{subsec:affected_variables}

Let $w$ index an attack window with labeled interval $[s_0,s_1)$ and length $L_{\text{lab}}=s_1-s_0$. A ramp length
\begin{equation}
L_{\text{ramp}} \triangleq \max\!\left(1,\left\lfloor \texttt{RAMP\_FRAC}\,L_{\text{lab}} \right\rceil \right)
\end{equation}
defines a piecewise-linear ramp $r_w(t)\in[0,1]$ over $t\in[s_0,s_1)$:
\begin{equation}
r_w(t)=
\begin{cases}
\displaystyle \frac{t-s_0+1}{L_{\text{ramp}}}, & s_0 \le t < s_0+L_{\text{ramp}},\\[6pt]
1, & s_0+L_{\text{ramp}} \le t < s_1.
\end{cases}
\end{equation}
All perturbations are multiplied by $r_w(t)$ and the activity gate $a_i(t)$.

\paragraph{Shadow-loss sampling and application.}
A window-specific shadow-loss $\Delta s_w$ (dB) is sampled from a truncated Gaussian mixture (parameters in \texttt{ATTACK\_SHADOW\_*}) and applied to the shadowing latent on active epochs:
\begin{equation}
s_{i,\mathrm{dB}}(t) \leftarrow s_{i,\mathrm{dB}}(t) - \Delta s_w\,r_w(t)\,a_i(t),
\qquad t\in[s_0,s_1),
\end{equation}
consistent with measured proximity/body-shadowing effects that induce excess attenuation without protocol manipulation \cite{januszkiewicz2018shadowing,wilson2010radio}. The baseline shadowing process itself follows standardized log-normal abstractions and decorrelation behavior \cite{3GPP38901}.

\paragraph{Coherence reduction and innovation increase.}
Let $h_i(t)\in\mathbb{C}$ denote the split-local complex fading process. In the normal regime, $h_i(t)$ follows a complex Gauss--Markov recursion with technology-conditioned correlation $\rho_i\in[0,1)$, a standard surrogate for temporally correlated small-scale fading \cite{goldsmith2005wireless,rappaport2010wireless}. For window $w$, a latent ``coherence-drop'' magnitude $k_w$ (dB) is sampled from a truncated mixture (parameters in \texttt{ATTACK\_KDROP\_*}) and mapped deterministically to (i) a correlation reduction factor $\delta_{\rho,w}\in(0,1]$ and (ii) an innovation multiplier $m_{\sigma,w}\ge 1$ (via \texttt{ATTACK\_ALPHA\_A0/A1} and \texttt{ATTACK\_SIGMA\_S0/S1}). A two-state Gilbert--Elliott excursion process over $[s_0,s_1)$ (parameters \texttt{ATTACK\_GE\_*}) introduces occasional worse conditions via a binary state $g_w(t)\in\{0,1\}$.

During the window, the attacked update uses a time-varying effective correlation $\rho_i(t)$ and innovation scaling $\nu_i(t)$ defined as
\[
\rho_i(t)\triangleq \rho_i\,\delta_{\rho,w}^{\,1+g_w(t)}, \qquad
\nu_i(t)\triangleq m_{\sigma,w}^{\,1+g_w(t)},
\]
so that the ``bad'' state ($g_w(t)=1$) yields a stronger coherence drop and larger innovation. On active epochs, the attacked update is generated as
\begin{align}
h_i^{\star}(t) &=
\rho_i(t)\,h_i(t\!-\!1)
+\sqrt{1-\rho_i(t)^2}\;\nu_i(t)\,w_i(t), \nonumber\\
w_i(t) &\sim \mathcal{CN}(0,1),
\end{align}
then blended using ramp-up and activity gating:
\begin{equation}
h_i(t) \leftarrow (1-r_w(t)a_i(t))\,h_i(t) + r_w(t)a_i(t)\,h_i^{\star}(t),
\end{equation}
where $r_w(t)\in[0,1]$ is the window ramp profile and $a_i(t)\triangleq\mathbb{I}\{\texttt{tx\_count}_i(t)>0\}$ is the activity gate. This construction targets the coherence/scattering changes commonly exploited in CSI-based presence detection \cite{palipana2016channel,zhou2015multipath,halperin2011tool}.

\paragraph{Optional Wi-Fi reflection component.}
For Wi-Fi nodes, an additional reflected path component can be injected during the window to emulate the appearance of a person-induced multipath component, using a Rician-amplitude surrogate (parameters \texttt{Wi-Fi\_REFLECT\_*}) \cite{palipana2016channel,zhou2015multipath}.

\paragraph{Unperturbed interference.}
The interference latent $i_{i,\mathrm{dB}}(t)$ is not modified by the attacker, consistent with a receive-only model. Its baseline burstiness (especially for PLC) is modeled using impulsive-noise abstractions widely used in PLC literature \cite{IEEE1901,middleton1977statistical,katayama2006noise,zimmermann2002multipath}.

\subsection{Coherent Re-Derivation and Window Manifest}
\label{subsec:coherent_rederivation}

After perturbing $h_i(t)$ and $s_{i,\mathrm{dB}}(t)$ on active epochs, all exported observables are recomputed deterministically using the same normal-regime mapping (including the same bounded measurement noise, quantization, and numerical clipping). This forces attack signatures to arise only through physically motivated dependencies rather than injected event flags, which is central for leak-safe benchmarking. Window definitions and sampled perturbation parameters are recorded in \texttt{attacks\_windows\_meta.csv} for reproducibility and controlled ablations.

\begin{figure}[t]
\centering
\includegraphics[width=\linewidth]{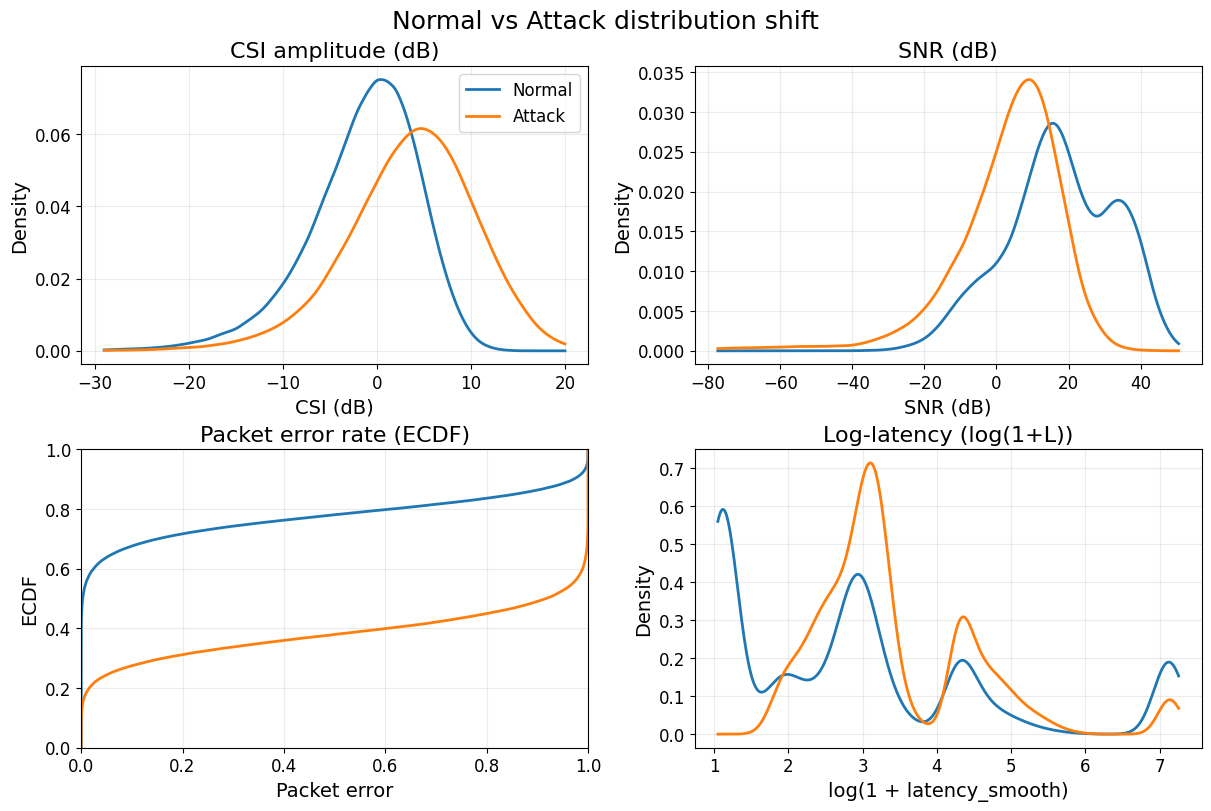}
\caption{Normal vs. attack distribution shift on \emph{active} samples (tx\_count$>0$). KDE overlays are shown for CSI (dB), SNR, and log-latency, while packet error is shown via ECDF (bounded support). Presence-only perturbations (shadow-loss and coherence degradation) shift CSI/SNR downward and propagate to higher packet error and latency through the coherent channel-to-metrics chain. Multi-modality reflects heterogeneous technologies (ZigBee/Wi-Fi/LTE/LoRa/PLC).}
\label{fig:dist_shift_active}
\end{figure}

\paragraph{Distribution-shift sanity check:}
To verify that presence-only perturbations propagate coherently through the exported chain, Fig.~\ref{fig:dist_shift_active} compares normal versus attack distributions on \emph{active} samples (tx\_count$>0$). Attacks induce a consistent shift in CSI amplitude (dB) and SNR, which increases packet error and propagates to higher delay under the ARQ-inspired latency mapping. Multi-modality reflects the heterogeneous technologies present in the HAN/NAN/WAN benchmark.

\section{Dataset Schema and Release}
\label{sec:stats_schema}

\subsection{Scale, Splits, and Node Inventory}
\label{subsec:scale_splits}

Let $N$ denote the number of nodes ($N=12$ in the released configuration). Each node is exported as three split-specific time series with lengths
$T_{\mathrm{train}}$, $T_{\mathrm{val}}$, and $T_{\mathrm{test}}$ (parameters \texttt{T\_TRAIN}, \texttt{T\_VAL}, \texttt{T\_TEST}), generated with an additional burn-in of $B$ epochs (\texttt{BURN\_IN}) that is not exported. Split realizations are independent (split-specific seeds), so latent state is not shared across train/validation/test.

Table~\ref{tab:node_inventory} summarizes node roles, tiers, technologies, and passive-attack eligibility. Eligibility is determined by membership in \texttt{ATTACK\_ELIGIBLE\_TECH}; backbone-wired media are excluded by construction.

\begin{table}[t]
\centering
\caption{Node inventory and passive-attack eligibility.}
\label{tab:node_inventory}
\scriptsize
\begin{tabular}{@{}c l c l c@{}}
\toprule
ID & Role & Tier & Comm.\ Tech & Eligible \\
\midrule
0  & Smart Meter    & HAN & ZigBee & Yes \\
1  & Smart Meter    & HAN & ZigBee & Yes \\
2  & Smart Meter    & HAN & Wi-Fi   & Yes \\
3  & Gateway        & HAN & Wi-Fi   & Yes \\
4  & DER            & NAN & LoRa   & Yes \\
5  & DER            & NAN & LoRa   & Yes \\
6  & Feeder Relay   & NAN & PLC    & Yes \\
7  & Controller     & NAN & LTE    & Yes \\
8  & PMU            & WAN & Fiber  & No  \\
9  & SCADA          & WAN & Fiber  & No  \\
10 & AMI Headend    & WAN & LTE    & Yes \\
11 & Substation GW  & WAN & PLC    & Yes \\
\bottomrule
\end{tabular}
\end{table}

\subsection{Coverage Control and Window Metadata}
\label{subsec:coverage_meta}

Attack prevalence is controlled by a configurable target fraction $r$ over \emph{active} rows of eligible nodes (parameter \texttt{TARGET\_ATTACK\_FRAC}). For eligible node $i$, define the activity indicator
$a_i(t)\triangleq \mathbb{I}[\texttt{tx\_count}_i(t)>0]$, the number of active rows
$A_i \triangleq \sum_{t} a_i(t)$, and the number of labeled attack rows
$Y_i \triangleq \sum_{t}\mathbb{I}[\texttt{attack\_label}_i(t)=1]$.
The realized active-row coverage is
\begin{equation}
r_i \triangleq \frac{Y_i}{\max(1,A_i)} .
\end{equation}
Window boundaries and window-level perturbation parameters (including the sampled \texttt{kdrop\_db} and \texttt{shadow\_loss\_db}, and the mapped \texttt{alpha\_drop} and \texttt{sigma\_mult}) are recorded in \texttt{attacks\_windows\_meta.csv} to support reproducibility and controlled ablations.

\section{Federated Baseline Detectors}
\label{sec:federated_baselines}

The primary objective of this work is the dataset generator and its leak-safe, propagation-consistent construction rather than proposing a new detector. Nevertheless, lightweight federated baselines are reported as a \emph{sanity check} to demonstrate that presence-only perturbations are (i) \emph{subtle}, (ii) \emph{technology dependent}, and (iii) not trivially separable without explicitly leveraging spatiotemporal structure and topology-aware context.

\subsection{Protocol and Feature Subset}
To keep the evaluation aligned with the threat model, baselines use a compact, leak-safe feature subset derived from measurable link indicators and topology-based neighbor summaries, while excluding diagnostic latents (\texttt{shadow\_db}, \texttt{interf\_db}) and activity variables as inputs. Specifically, the subset contains standardized (train-only, per-node) versions of: local link indicators (\texttt{SNR}, \texttt{C}, \texttt{PER}, \texttt{$\tilde{L}$}), phase/coherence surrogates (\texttt{phase\_sin}, \texttt{phase\_cos}, \texttt{dphase}), and neighbor-averaged counterparts (\texttt{avg\_neighbor\_*}) computed via the mixing matrix in~\eqref{sec:topology_features}. To avoid inflation from silent intervals, training and evaluation are restricted to \emph{active} rows (\texttt{tx\_count}$>0$). Fiber-only clients (PMU$_8$, SCADA$_9$) are excluded, consistent with the attack-eligibility policy.

\subsection{Row-wise Federated Baselines (No Temporal Modeling)}
The reported results intentionally use a \emph{row-wise} setting (no temporal windows) to establish a lower-bound baseline when only per-sample observables and limited neighbor context are available.

Table~\ref{tab:rowwise_fed_baselines} summarizes macro-averaged test performance over the $10$ non-fiber clients. The linear baseline (Fed-LR) attains high recall but limited precision, reflecting frequent false alarms when temporal consistency is ignored. Tree ensembles (Fed-XGB) improve precision but can miss attacks in the more subtle regimes. Recurrent baselines evaluated in the same row-wise regime (Fed-LSTM with sequence length $1$; Fed-GRNN as a gated recurrent baseline with sequence length $1$) yield intermediate behavior. Overall, these tradeoffs support the intended benchmark design: presence-only effects manifest as low-amplitude, correlated shifts that motivate detectors that incorporate \emph{time} and \emph{topology-aware} context beyond per-sample decisions.

\begin{table}[t]
\centering
\caption{Row-wise federated baselines on non-fiber clients. These results are included to demonstrate dataset subtlety and heterogeneity rather than to optimize detection performance.}
\label{tab:rowwise_fed_baselines}
\scriptsize
\begin{tabular}{@{}lcccc@{}}
\toprule
Model & Precision & Recall & F1 & Accuracy \\
\midrule
Fed-LR   & 0.3997 & 0.8866 & 0.5326 & 0.7301 \\
Fed-XGB  & 0.5469 & 0.6634 & 0.7129 & 0.8192 \\
Fed-LSTM & 0.5793 & 0.7788 & 0.6489 & 0.8580 \\
Fed-GRNN & 0.6813 & 0.7857 & 0.7201 & 0.8954 \\
\bottomrule
\end{tabular}
\end{table}

\subsection{Interpretation and Implication for Spatiotemporal Pipelines}
Two observations follow directly from Table~\ref{tab:rowwise_fed_baselines}. First, performance varies across technologies and clients (e.g., low-prevalence LoRa clients are noticeably harder), consistent with the generator’s tier/technology-conditioned channel and impairment dynamics. Second, the precision--recall tradeoffs indicate that per-epoch decisions are not consistently robust for stealthy presence-only perturbations. This motivates the use of graph-temporal pipelines that combine (i) temporal consistency across epochs and (ii) spatial context across adjacent nodes, which is supported by the dataset release (per-node splits, adjacency, neighbor aggregates, and stored train-only normalization).

\section{Conclusion}
This paper introduced an IEEE-inspired, literature-anchored benchmark dataset generator for presence-only passive reconnaissance in tiered smart-grid communications. The benchmark targets a receive-only adversary whose impact arises solely through proximity-induced propagation perturbations, modeled as windowed excess shadowing and coherence degradation with increased innovation. A key design objective was physical and statistical consistency: exported observables are obtained through a deterministic channel-to-metrics mapping in which measurement-realistic CSI proxies induce SNR shifts that propagate to packet error behavior and delay dynamics, without injected event flags or protocol-layer shortcuts. The release further enforces leak-safe evaluation via split-independent realizations with burn-in removal, strictly causal temporal descriptors, and train-only per-node standardization applied to validation/test using stored parameters, alongside tier-aware adjacency and neighbor-context features to support topology-aware learning.

Baseline federated detectors demonstrated non-trivial precision--recall tradeoffs under heterogeneous technologies, confirming that the perturbations are subtle and not uniformly separable from single-epoch statistics. The dataset and accompanying artifacts therefore enable standardized, reproducible comparisons of centralized, local, and federated graph-temporal pipelines for passive reconnaissance detection in smart-grid communication graphs.

\section*{Data and Code Availability}
The synthetic dataset generator and scripts to reproduce the results are publicly available at
\url{https://github.com/bochraagha/smartgrid-passive-attack-dataset-generator}.

\section*{Acknowledgment}
The authors thank American University of Beirut University Research Board for supporting this
work and for providing the research environment that enabled the dataset development and benchmarking.

A large language model (LLM) was used only for minor language editing to improve clarity and grammar. All technical content, analysis, and conclusions were produced and verified by the authors.

\bibliographystyle{IEEEtran}
\bibliography{references}

\end{document}